\documentclass[twocolumn,prl,aps,floatfix,superscriptaddress,showpacs]{revtex4}
\usepackage{graphicx}
\usepackage{dcolumn}
\usepackage{bm}

\begin{document}
\title{Polarized resonant inelastic x-ray scattering as an 
ultra-fine probe of excited states in La$_2$CuO$_4$}

\author{Abhay Shukla},
\affiliation{Institut de Min\'eralogie et de la Physique des Milieux Condens\'ees,
4 Place Jussieu, 75252, Paris cedex 05, France}
\author{Matteo Calandra}
\affiliation{Institut de Min\'eralogie et de la Physique des Milieux Condens\'ees,
4 Place Jussieu, 75252, Paris cedex 05, France}
\author{Munetaka Taguchi}
\affiliation{Soft X-Ray spectroscopy Lab, RIKEN/SPring-8, 1-1-1, Mikazuki, Sayo, Hyogo 679-5148, Japan}
\author{Akio Kotani}
\affiliation{Soft X-Ray spectroscopy Lab, RIKEN/SPring-8, 1-1-1, Mikazuki, Sayo, Hyogo 679-5148, Japan}
\affiliation{Photon Factory,IMSS, High Energy Accelerator Research Organisation, Tsukuba, Ibaraki 305-0801 Japan}
\author{G. Vanko}
\affiliation{European Synchrotron Radiation Facility, BP 220, F-38043 Grenoble cedex, France}
\author{S-W. Cheong}
\affiliation{Department of Physics and Astronomy, Rutgers University, Piscataway, New Jersey 08854, USA}       

\date{\today}

\begin{abstract}
X-ray absorption is the standard method to probe the unoccupied density of states
at a given edge. Here we show that polarized Resonant Inelastic X-Ray Scattering
in La$_2$CuO$_4$ at the Cu K-edge is extremely sensitive to the environment of the Cu atom and
the fine structure in the Cu 4p density of states. 
Combined {\it ab initio} and many-body cluster calculations, used for the first time in such a context, 
show remarkable agreement with experiment. In particular we identify 
a non-local effect namely a transition to {\it off-site} Cu 3d states.

\end{abstract}

\maketitle

X-ray absorption spectroscopy (XAS) has long been used to investigate the nature of the 
partial density of excited states at a given absorption threshold. 
However pre-edge features, mainly composed of quadrupolar excitations,
are often hard to detect in XAS due to their small intensity and the broadening due to the core-hole lifetime.
As a consequence
important insights related to the local environment of the absorbing atom
are  not accessible in XAS. Here we show that resonant inelastic X-ray 
scattering (RIXS) yields a precise determination of the 
low-energy unoccupied electronic states and the chemical environment of the absorbing atom.
RIXS is a  powerful tool for the study of the electronic excitations 
in solids\cite{KotaniReview}
giving access to site, element and orbital selective information.
It is a second order process where the X-ray emission is measured as the incident 
X-ray energy is tuned through an absorption edge of the sample. The incident
energy determines the intermediate state and the core-hole enhances the cross section 
\cite{Kao}.

In this work we study resonant Cu 1s-2p emission in La$_2$CuO$_4$
both from an experimental and theoretical point of view.
The excited states in La$_2$CuO$_4$ are of interest because of the correlated nature of this
compound which on hole doping becomes a high temperature superconductor. 
The hopping between the Cu $3d_{x^2-y^2}$ and the O $2p$ orbitals
combined with Coulomb repulsion between Cu 3d electrons form the basis 
for describing the 
low energy excitations in this solid \cite{ZhangRice}.
In the intermediate state of the resonant process a 1s Cu core-electron is excited 
to states with energies around the pre-edge region of the Cu K-absorption edge. 
The final state involves K$\alpha$ emission with filling of the 1s hole at the expense of a hole
in the 2p state.  Identifying the measured features 
and explaining their dependence on incident energy and incident photon polarisation
is not a trivial task from the theoretical point of view,
particularly in strongly correlated 
systems where single particle theories like standard density functional theory  (DFT)
dramatically fail due to the lack of important 
many-body effects. A different approach to the problem
is the use of many-body cluster models based on multiband Hubbard-like hamiltonian.
While these methods correctly include many-body effects they often require 
some knowledge of the empty state density of states (DOS) which cannot be 
easily obtained without the use of {\it ab initio} electronic structure calculations.
For this reason core level spectroscopy is an ideal field for the application of combined
{\it ab initio} and many-body cluster model approaches.

\begin{figure}
\includegraphics[scale=.35]{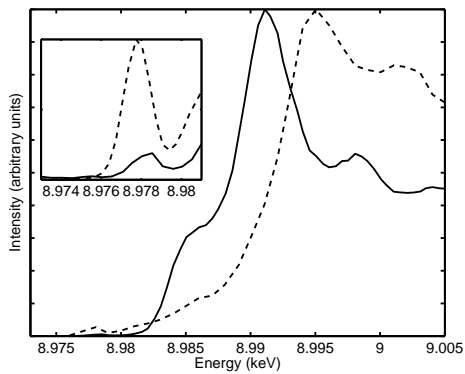}
\caption{\label{figpfy}     
Partial fluorescence yield absorption spectra for La$_2$CuO$_4$ in the a-b geometry (dashed line)
showing transitions to Cu 4p$_{\sigma}$ orbitals and the c geometry (solid line) showing transitions to Cu 4p$_\pi$ orbitals. 
Each spectrum is normalized to its peak. The inset is a zoom of the pre-edge area where resonant scans are taken.
 }
\end{figure}

\begin{figure}
\includegraphics[scale=.35]{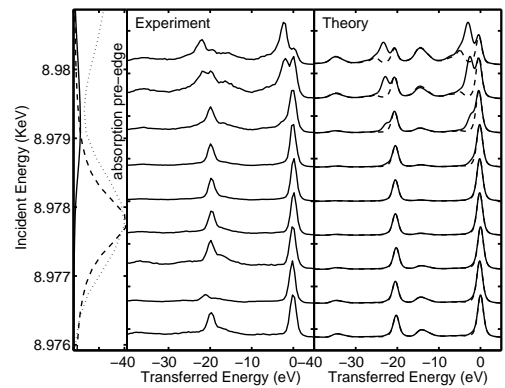}
\caption{\label{figab}     
{\it a/b} geometry. Left panel: The pre-edge region of the absorption spectrum. Experiment (dotted line), 
cluster-model calculated {\it on-site} Cu 1s-3d quadrupolar (dashed line) and {\it off-site} Cu 1s-3d dipolar contributions (solid line).
Center panel: Measured RIXS spectra. The incident energy can be
traced to the absorption curve in the left panel. Right panel: Calculated RIXS spectra using {\it ab initio}
Cu 4p DOS (solid line) and the modified DOS  shown in the inset of Fig. \ref{figDOS} (dashed line). The peak at 0 eV 
is due to the {\it on-site} Cu 1s-3d transition. The additional peak appearing at -3 eV
in the RIXS spectra is due to the lowest energy feature of the 4p$_{\sigma}$ DOS
which is determined by the hybridization with {\it off-site} Cu 3d empty states.
}
\end{figure}

\begin{figure}
\includegraphics[scale=.35]{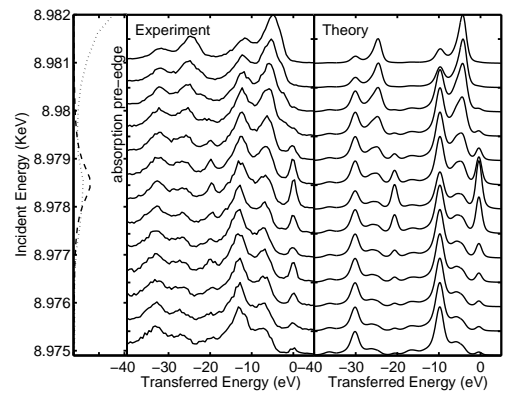}
\caption{\label{figc}     
{\it c} geometry. Left panel: The pre-edge region of the absorption spectrum. Experiment (dotted line), 
calculated {\it on-site} Cu 1s-3d quadrupolar contribution (dashed line).
Center panel: Measured RIXS spectra. The incident energy can be
traced to the absorption curve in the left panel. Right panel: Calculated RIXS spectra.
 }
\end{figure}

The orthorhombic structure of La$_2$CuO$_4$ with the CuO planes oriented 
perpendicular to the c-axis allows the incident X-ray polarization to be 
unequivocally defined. Our twinned La$_2$CuO$_4$ single crystal was a plate-like
rectangular sample with the c-axis perpendicular to the sample plane and the 
a/b axis parallel to the sides of the sample.
The experiment was performed at beamline ID16 at
the European Synchrotron Radiation Facility. The
experimental setup was that of a Rowland circle spectrometer with a horizontal
scattering plane, using a spherically focusing, bent Si(444) analyzer \cite{Shukla}.
The absorption spectrum of La$_2$CuO$_4$ was obtained by measuring the total 
fluorescence yield. We also measured the partial 
fluorescence yield spectrum by using the spectrometer set to the peak of the 
$K\alpha_{1}$ fluorescence line and varying the incident energy as for an
absorption measurement. These were measured for two orientations of the sample: 
In the first geometry (noted {\it a/b} henceforth) the a-b plane was in the horizontal 
scattering plane such that it contained both the incident photon momentum and 
the polarization. They were both oriented at $45^\circ$ with respect to the a/b axis.
In the second geometry (noted {\it c}) the a-b plane, oriented vertically, was perpendicular to 
the scattering plane. The polarization was
at $ 75^\circ$ with respect to the a-b plane ($15^\circ$ with respect to the c-axis of the sample). These geometries were also 
used for the measurement of the 1s2p resonant emission. It should be noted that 
due to the in-plane polarization of the photons the
 {\it a/b} geometry is sensitive to the in-plane or 
$\sigma$ orbitals only while the {\it c} geometry is sensitive primarily to the $\pi$ orbitals.
The absorption spectrum is characterized by a small pre-peak at the onset, 
usually attributed at least partly to {\it on-site} Cu 1s-3d quadrupolar transitions. 
In Fig. \ref{figpfy} we show the partial fluorescence yield spectra 
for both geometries. 
In our RIXS measurements the incident energy is limited to 
the region around the pre-peak (inset of Fig. \ref{figpfy} and left panels of Fig. \ref{figab} and Fig. \ref{figc}). 
It will be seen that the dipolar contribution in this 
pre-edge region being small, the weak {\it on-site} Cu 1s-3d quadrupolar transition is easily visible.
Most importantly we will show that an even weaker {\it off-site} Cu 1s-3d dipolar transition mediated by 
O2p states and never before identified is also clearly visible
in the RIXS spectra. Though the absorption spectrum is relatively featureless, 
the emission spectra are rich in structure and strongly 
dependent on the incident energy.

In Fig. \ref{figab} and Fig. \ref{figc} (center panels) we show the experimental RIXS spectra for 
incident energies increasing from bottom up.
The baseline of each RIXS spectrum corresponds to the incident energy at which it was measured
as seen on the absorption spectrum on the left panel. RIXS spectra have been arbitrarily shifted such that the 
lowest  energy transfer feature corresponds to
the zero of the energy scale and have been normalized to the most intense feature in each spectrum.
They show twin features separated by 20 eV due to spin-orbit splitting of the 2p core level. In the following we shall discuss
only the region from 0 eV to -20eV energy transfer. The spectra are 
dominated by a sharp peak at 0 eV which resonates when excitation energies are on the absorption pre-peak. 
In CuO a similar structure has been interpreted as due to a quadrupolar transition\cite{Hayashi,Doring}. 
By separately calculating the contribution of $1s\to 3d$ quadrupolar,
and $1s\to 4p_{\sigma}$ and $1s\to 4p_{\pi}$ dipolar transitions we easily confirm that in 
La$_2$CuO$_4$ this peak is due to the quadrupolar transition.
In both geometries two additional features are seen.
In the $a/b$ geometry (Fig.\ \ref{figab}) a weak, broad, peak is observed around -17 eV 
while a sharp peak at about -3 eV resonates only at higher incident energies. 
In the $c$ geometry (Fig.\ \ref{figc}) a strong broad structure with twin peaks at -12 eV and -7 eV is seen while a second peak  
appears at -5 eV, resonating at higher incident energies.
The calculated results of the Cu K$\alpha$ RIXS are shown in Fig.\ \ref{figab} and Fig.\ \ref{figc} (right panels) and are
in excellent agreement with the experimental ones. The broad structures at higher energy loss are attributed to virtual transitions due
to the main peaks in the Cu 4p DOS and are slightly shifted in energy from the measured ones. 
The lower energy resonant structures are due to transitions to features in the tails of the DOS. 
These attributions were made by modifying portions of the calculated DOS and tracing the modifications in the calculated RIXS spectrum.

RIXS data are interpreted using a unified {\it ab initio} and multiplet calculations 
approach. Numerical multiplet 
 calculations were carried out based on CuO$_{6}^{10-}$ 
single cluster model with intra-atomic full multiplets in D$_{4h}$ local symmetry and including the screening channel 
for charge transfer from O $2p$ ligand state. 
Only a single Cu atom is retained and charge transfer between 
the Cu 3d state and the neighboring O 2p orbitals is allowed. 
The Hamiltonian is given by 
$H=H_0+H_{pd}
+H_{ch}+H_{\rm multiplet}$. $H_0$ includes the on-site energies of the 
Cu 1s ($\varepsilon_{1s}$), Cu 2p ($\varepsilon_{2p}$), 
Cu 3d ($\varepsilon_{3d}(\Gamma)$,), Cu 4p ($\varepsilon_{4p}(k)$) and
Ligand O 2p ($\varepsilon_{p}(\Gamma)$) states;
$H_0=\sum_{\Gamma,\sigma}\varepsilon_{3d}(\Gamma)d^{\dagger}_{\Gamma\sigma}d_{\Gamma\sigma}
+ \sum_{m,\sigma}\varepsilon_{2p}p^{\dagger}_{m \sigma}p_{m \sigma}
 + \sum_{\sigma}\varepsilon_{1s}s^{\dagger}_{\sigma}s_{\sigma}
+ \sum_{k,\sigma}\varepsilon_{4p}(k)P^{\dagger}_{k\sigma}P_{k \sigma} 
+\sum_{\Gamma,\sigma}\varepsilon_{p}(\Gamma)a^{\dagger}_{\Gamma\sigma}a_{\Gamma\sigma} $.
The irreducible 
representations of the D$_{4h}$ symmetry are 
$\Gamma$ ( = $b_{1g}$, $a_{1g}$, $b_{2g}$ and $e_{g}$).
$H_{pd}$ is the hybridization between Cu 3d and O 2p ligand states,
$H_{pd}= \sum_{\Gamma,\sigma}V(\Gamma)(d^{\dagger}_{\Gamma\sigma}a_{\Gamma\sigma}
 + a^{\dagger}_{\Gamma\sigma}d_{\Gamma\sigma})$.
The core-hole
interaction $H_{dc}$ is composed of two contributions, 
the attraction between Cu 3d electrons and Cu 1s holes ($U_{dc}(1s)$) and
between Cu 3d electrons and Cu 2p holes, ($U_{dc}(2p)$), namely
$H_{dc}= U_{dc}(1s)\sum_{\Gamma,\sigma,\sigma'}d^{\dagger}_{\Gamma\sigma}d_{\Gamma \sigma}
(1 - s^{\dagger}_{\sigma'}s_{\sigma'}) 
 + U_{dc}(2p)\sum_{\Gamma,m,\sigma,\sigma'}d^{\dagger}_{\Gamma\sigma}d_{\Gamma\sigma}
(1 - p^{\dagger}_{m\sigma'}p_{m\sigma'})$. Finally
the Hamiltonian $H_{\rm{multiplet}}$ describes the intra-atomic multiplet coupling 
originating from the multipole components of the Coulomb interaction 
between Cu 3$d$ states and that between Cu 3$d$ and 2$p$ or 1$s$ states and 
the spin-orbit interactions for Cu 3$d$ and 2$p$ states \cite{cow81,esk90,tag00,sho03}.

The parameter values used are (in eV): 
$U_{dc}(1s)=U_{dc}(2p)=-7.0$, $V(b_{1g})=2.5$, $V(a_{1g})=1.44$, $V(b_{2g})=1.25$, 
$V(e_{g})=0.88$, $\Delta=2.75$(charge transfer energy), $T_{pp}=1.0$ (hybridization between nearest neighbor O $2p$ orbitals),
very close to the ones used in ref. \cite{esk90}. 
The Slater integrals and the spin-orbit coupling constants are calculated by 
Cowan's Hartree-Fock program,~\cite{cow81} 
and then the Slater integrals are scaled down to 80{\%}. 

The K$\alpha$ RIXS cross-section is calculated on the basis of the formula of 
the coherent second order optical process as 
$S(\Omega,\omega)=S_D(\Omega,\omega)+S_Q(\Omega,\omega)$ where $S_{\alpha}(\Omega,\omega)$ 
with $\alpha=D,Q$ is:
\begin{eqnarray}
S_{\alpha}(\Omega,\omega) &=&
 \sum_{f} \biggl|\sum_{m} \frac{ \langle f \mid T_{D} \mid m \rangle \langle m 
\mid T_{\alpha} \mid g \rangle}{E_{g}+ \Omega-E_{m}-i\Gamma_{K}}\biggr|^{2} \nonumber \\
&\times&\frac{\Gamma_{L}/\pi}
{(E_{g}+ \Omega - E_{f}- \omega)^{2} +\Gamma_{L}^{2}} \ \ ,
\label{eq:RIXS}
\end{eqnarray} 
and $\mid g \rangle$, $\mid m \rangle$ and $\mid f \rangle$ are the ground, 
intermediate and final states of the Hamiltonian $H$ with energies $E_g, E_m$ and $E_f$, respectively. 
The incident and emitted photon energies are represented by $\Omega$ and $\omega$, respectively. 
The core-hole lifetime broadening is denoted by $\Gamma_K=0.4$ eV, half width at half maximum (HWHM) for the 1$s$ core-hole and 
$\Gamma_L=0.2$ eV (HWHM) for the 2$p$ core-hole. 
Gaussian broadening due to the experimental resolution is taken to be 0.6 eV (HWHM).  
The operators $T_{\alpha}$ with $\alpha=D,Q$ represent the optical dipole and quadrupole 
transition respectively. The polarization of the incident photon is also taken into account \cite{tag00,sho03}.

To calculate eq. \ref{eq:RIXS}, we use, as basis states, 
two configurations, $3d^{9}$ and $3d^{10}\underline{L}$, 
where $\underline{L}$ represents a hole in the ligand states. 
The basis configurations in the intermediate state $\mid m \rangle$ are 
$\underline{1s}3d^{10}$ for the quadrupole transition 
and $\underline{1s}3d^9 4p$ and $\underline{1s}3d^{10} \underline{L}4p$ 
for the dipole excitation. The final states $\mid f \rangle$ are $\underline{2p}3d^{10}$  (quadrupole) 
and $\underline{2p}3d^9 4p$ and $\underline{2p}3d^{10} \underline{L}4p$  (dipole).
In the case of dipolar transitions we assume the intermediate states $\mid m \rangle$
(final states $\mid f \rangle$) 
to be a direct product of a Cu $4p$ state $\mid k \rangle$ and the remaining system 
$\mid m' \rangle$ ( $\mid f' \rangle = \mid k \rangle \mid m' \rangle$) with energies 
$E_m= E_{m'} + \varepsilon_k$ ($E_f= E_{f'} + \varepsilon_k$). 
Thus we can transform the sum over $m$ in eq. \ref{eq:RIXS} to a sum over $m'$ plus an integral over
the high enenergy continuum weighted over the 4p Cu projected density of states $\rho_{\eta}(\epsilon)$
(with $\eta=\sigma,\pi$),
namely: 
$\sum_{m}\longmapsto \sum_{m'} \int d\epsilon \rho_{\eta}(\epsilon)$\cite{tag00}.

\begin{figure}
\includegraphics[scale=.35]{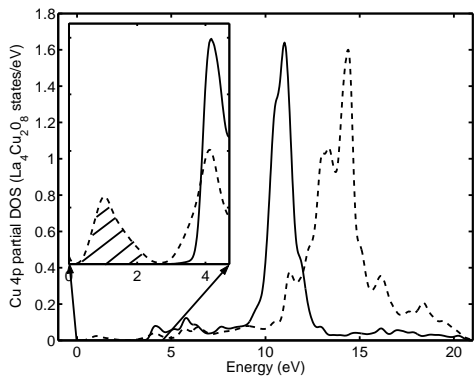}
\caption{\label{figDOS}     
Projected Cu $4p_{\pi}$ (solid) and $4p_{\sigma}$ (dashed line) orbital 
density of states calculated {\it ab initio} . The inset shows the low energy region in detail.
The hatched peak, the lowest energy structure in the Cu 4p$_{\sigma}$ DOS, is induced by hydridization
between the Cu 4p$_{x,y}$ atomic orbitals, the empty O 2p orbitals and the empty 3d orbitals of the neighbouring Cu atoms
forming a $b_{1g}$ state.}
\end{figure}

The Cu 4p projected  density of states, 
$\rho_{\eta}(\epsilon)=
\frac{1}{N_k}\sum_i \sum_{kn}|\langle\psi|{\rm Cu^{i}}  4p_{\eta}\rangle|^2\delta(\epsilon_{{\bf k}n}-\epsilon)$, 
is calculated {\it ab initio}\cite{pwscf} using
the Kohn-Sham eigenvalues ($\epsilon_{{\bf k}n}$) and eigenfunctions $(\psi_{{\bf k}n})$
and the $|{\rm Cu^{i}} 4p_{\eta}\rangle$
orthonormalized atomic orbitals.
We disregard the core hole potential acting on the 4p state,
because it is screened strongly by a 3d electron
transferred from the O 2p states. 
Since we consider a supercell with two La$_{2}$CuO$_{4}$ molecules,
$\rho_{\eta}$ is also summed over the two inequivalent Cu atoms ($i=1,2$). 
In the {\it ab initio} simulations we use
ultrasoft pseudopotentials \cite{Vanderbilt} with a 45 Ry cutoff and  
the spin-polarized generalized gradient approximation(SPGGA) \cite{PBE} 
for the exchange-correlation kernel. Since using the SPGGA the ground state
turns out to be metallic we
adopted the SPGGA+U approximation \cite{Anisimov} 
with $U=8 \,$eV in the implementation of 
ref. \cite{Cococcioni}. 
Self-consistence is performed
over a $6\times 6\times 6$ k-points mesh in the Brillouin zone centered at
the  $\Gamma$ point. 
We converge to an antiferromagnetic
state with Cu local magnetic moment $ 0.625 \mu_{B}$, to be compared with the 
experimental value of $0.55\sim0.60 \mu_{B}$\cite{Czyzyk,Endoh}.
The band gap is approximately $0.65$ eV, 
in good agreement with previous calculations
of ref. \cite{Nekrasov} (0.7 eV), but substantially smaller than the one
determined from optical conductivity data in ref. \cite{Uchida} ($\approx 2$ eV).
The Cu 4p$_{\sigma,\pi}$ projected 
DOS calculated on a $N_k=20\times 20\times 20$ k-points mesh 
are shown in Fig. \ref{figDOS}.

For a finer comparison between theory and experiment let us first
consider the sharp peak at 0 eV in Fig. \ref{figab} and Fig. \ref{figc} due to the on-site 1s-3d quadrupolar excitations, 
where the intermediate state is $\underline{1s}3d^{10}$ 
and the final state is $\underline{2p}3d^{10}$. The $\underline{2p}3d^{10}$ final state 
shows no multiplet structure because the 3d shell is filled, which explains the 
single peak spectral shape. 
Next, we consider the dipolar contribution.  In the $a/b$ geometry the polarization being in the a-b plane, 
only $1s \to 4p_{\sigma}$ transitions contribute. In the $c$ geometry $\pi$ transitions are visible,
the $\sigma$ component being almost negligible (see Fig. \ref{figc}). 
Finally we illustrate the extreme sensitivity of this method by looking for the origin
of the peak at -3 eV in the $a/b$ geometry. A closer look at the Cu 4p DOS in Fig. \ref{figDOS} reveals that though the main peak of 
the  $\sigma$ component is at higher energy than that of the $\pi$ component, the lowest energy structure is $\sigma$ in nature. 
The intensity of this structure, peaking at 1eV, is less than one percent of the total DOS
in this region. Nevertheless if this structure is artificially truncated (indicated by the hatching in Fig. \ref{figDOS}), 
the peak at -3 eV disappears in the calculated RIXS
(Fig.\ \ref{figab}, right panel, dashed curve).  This low energy structure in the Cu 4p$_{\sigma}$ DOS is induced by the
hybridization with O 2p states and with the Cu 3d $b_{1g}$ empty state of the neighbouring Cu sites. 
This is the first identification of this 1s-3d off-site transition mediated by 
the O 2p orbitals and shows the chemical detail which such a measurement reveals. It is also present in the calculated
absorption in the $a/b$ geometry (Fig. \ref{figab}, left panel) while absent in the calculated
absorption in the $c$ geometry (Fig. \ref{figc}, left panel) and in the  measured and calculated RIXS (Fig. \ref{figc}, center and right panels)

XANES measurements have long been the standard method for accessing excited states. 
Earlier works \cite{Krisch,Doring} have pointed to the advantages of RIXS. 
We have conclusively shown with the example of La$_2$CuO$_4$ that RIXS at the Cu K$\alpha$ 
line provides a sensitive and precise way to determine the Cu 4p unoccupied DOS and the interactions
of the Cu atom with its immediate environment. We distinguish between the dipolar and quadrupolar part of
the transitions and separate them according to their symmetry.
We also find that a new approach combining {\it ab initio} calculations of the partial Cu-4p DOS with 
many body cluster calculations is very successful in explaining the data. Indeed both the polarization 
and incident energy dependance of the spectra are fully explained. Features in the measured 
spectra are traced to fine details of the calculated Cu-4p DOS (used directly) giving a very sensitive method 
for checking its reliability and identifying for the first time the off-site Cu 1s-3d dipolar transition. 
In particular this work, by validating the nature of the Cu 4p 
unoccupied DOS, will lead to the definitive attribution of the origin of the different features 
in the Cu K-edge X-ray absorption spectrum \cite{Tolentino, Kosugi}. 
Finally this method should prove useful for investigating the change in 
these excited states produced by doping as well as those induced by changes in local symmetry, 
notably those between the Lanthanum based hole-doped 
compounds and Neodymium based electron doped compounds.
M.C. acknowledges discussions with G. A. Sawatzky. This work benefitted from
CNRS/JSPS research grant No. 16030.
Calculations were performed at the IDRIS supercomputing center 
(grant no. 051202). SWC was supported by the NSF-DMR-0405682.

\end{document}